\documentclass[12pt,twoside]{article}  
\usepackage{cite} 

\catcode `\@=11
\@addtoreset{equation}{section}

\def\section{\@startsection {section}{1}{\z@}{-3.5ex plus -1ex minus
     -.2ex}{2.3ex plus .2ex}{\normalsize\bf}}
\def\subsection{\@startsection{subsection}{2}{\z@}{-3.25ex plus -1ex minus
 -.2ex}{1.5ex plus .2ex}{\normalsize\bf}}

\def\thebibliography#1{\section*{{
				   \rm REFERENCES}\@mkboth
  {REFERENCES}{REFERENCES}}\list
  {[\arabic{enumi}]}{\settowidth\labelwidth{[#1]}\leftmargin\labelwidth
  \advance\leftmargin\labelsep
  \usecounter{enumi}}
  \def\newblock{\hskip .11em plus .33em minus -.07em}
  \sloppy
  \sfcode`\.=1000\relax}
 
\catcode `\@=12

\title{\vspace*{1.5cm} \normalsize\bf  
TOWARDS  
A ``PRE-CANONICAL'' 
QUANTIZATION OF 
  GRAVITY 
WITHOUT THE SPACE+TIME DECOMPOSITION  
}

\author{\vspace{1.5cm} \sc Igor V. Kanatchikov\thanks{e-mail: 
{\tt ikanat@ippt.gov.pl, kai@fuw.edu.pl}} \vspace{-1.5cm}\\    
\small Laboratory of Analytical Mechanics and Field Theory \vspace{-0.2cm}\\ 
\small Institute of Fundamental Technological Research\vspace{-0.2cm} \\
\small Polish Academy of Sciences\vspace{-0.2cm}\\
\small \'Swi\c etokrzyska 21,  Warsaw  PL-00-049, Poland \vspace{-0.2cm} \\
\small \centerline{\small and} \vspace{-0.2cm} \\ 
\small Theoretisch-Physikalisches Institut \vspace{-0.2cm} \\ 
\small Friedrich Schiller Universit\"at Jena \vspace{-0.2cm} \\ 
\small Max-Wien-Platz 1, Jena D-07743, Germany  
}

\date{}

\begin{document}

\maketitle 


\vspace*{-96mm}\vspace*{-2mm}
\hbox to 6.2truein{
\footnotesize\it 
\hfil \hbox to 0 truecm{\hss 
\normalsize\rm 
{\sf  FSU TPI \, 01/99} { }}\vspace*{-4mm}}
\hbox to 6.2truein{
\vspace*{-1mm}\footnotesize 
\hfil 
} 
\hbox to 6.2truein{
\vspace*{0mm}\footnotesize 
\hfil \hbox to 0 truecm{ 
\hss \normalsize 
\sf March 1999 
\hfil
}
} 
\hbox to 6.2truein{
  \footnotesize 
\hfil \hbox to 0 truecm{ 
\hss \normalsize 
\sf gr-qc/9909032  
}  
}

\vspace*{72mm} \vspace*{16mm}


\begin{flushright} 
\begin{minipage}{5.3in}
{\footnotesize 
{\footnotesize \bf Abstract.}  
Quantization of gravity is discussed in the context of 
field quantization based on 
 an 
 analogue of canonical formalism  
(the 
 De Donder-Weyl 
 canonical 
theory)
which does not require the space+time decomposition.  
 Using Ho\v{r}ava's (1991) De Donder-Weyl formulation of 
 General Relativity  
we 
arrive to  
a covariant   
generalization of the 
Schr\"odinger equation  
for the wave {\em function} 
 of 
space-time and metric  
variables 
 and a supplementary   
 ``bootstrap condition''    
which  
  self-consistently incorporates 
 the 
classical background space-time geometry 
  as a quantum average  
 and closes the system of equations. 
 Some open 
 questions for further research are outlined. \\

{\sl PACS: } {0460, 0420F, 0370, 1110E} 


 } 
\end{minipage}
\end{flushright} 


\newcommand{\beq}{\begin{equation}}
\newcommand{\eeq}{\end{equation}}
\newcommand{\beqa}{\begin{eqnarray}}
\newcommand{\eeqa}{\end{eqnarray}}
\newcommand{\nn}{\nonumber}

\newcommand{\half}{\frac{1}{2}}

\newcommand{\xt}{\tilde{X}}

\newcommand{\uind}[2]{^{#1_1 \, ... \, #1_{#2}} }
\newcommand{\lind}[2]{_{#1_1 \, ... \, #1_{#2}} }
\newcommand{\com}[2]{[#1,#2]_{-}} 
\newcommand{\acom}[2]{[#1,#2]_{+}} 
\newcommand{\compm}[2]{[#1,#2]_{\pm}}

\newcommand{\lie}[1]{\pounds_{#1}}
\newcommand{\co}{\circ}
\newcommand{\sgn}[1]{(-1)^{#1}}
\newcommand{\lbr}[2]{ [ \hspace*{-1.5pt} [ #1 , #2 ] \hspace*{-1.5pt} ] }
\newcommand{\lbrpm}[2]{ [ \hspace*{-1.5pt} [ #1 , #2 ] \hspace*{-1.5pt}
 ]_{\pm} }
\newcommand{\lbrp}[2]{ [ \hspace*{-1.5pt} [ #1 , #2 ] \hspace*{-1.5pt} ]_+ }
\newcommand{\lbrm}[2]{ [ \hspace*{-1.5pt} [ #1 , #2 ] \hspace*{-1.5pt} ]_- }
\newcommand{\pbr}[2]{ \{ \hspace*{-2.2pt} [ #1 , #2 ] \hspace*{-2.55pt} \} }
\newcommand{\we}{\wedge}
\newcommand{\dv}{d^V}
\newcommand{\nbrpq}[2]{\nbr{\xxi{#1}{1}}{\xxi{#2}{2}}}
\newcommand{\lieni}[2]{$\pounds$${}_{\stackrel{#1}{X}_{#2}}$  }

\newcommand{\rbox}[2]{\raisebox{#1}{#2}}
\newcommand{\xx}[1]{\raisebox{1pt}{$\stackrel{#1}{X}$}}
\newcommand{\xxi}[2]{\raisebox{1pt}{$\stackrel{#1}{X}$$_{#2}$}}
\newcommand{\ff}[1]{\raisebox{1pt}{$\stackrel{#1}{F}$}}
\newcommand{\dd}[1]{\raisebox{1pt}{$\stackrel{#1}{D}$}}
\newcommand{\nbr}[2]{{\bf[}#1 , #2{\bf ]}}
\newcommand{\der}{\partial}
\newcommand{\oo}{$\Omega$}
\newcommand{\Om}{\Omega}
\newcommand{\om}{\omega}
\newcommand{\eps}{\epsilon}
\newcommand{\si}{\sigma}
\newcommand{\Lm}{\bigwedge^*}

\newcommand{\inn}{\hspace*{2pt}\raisebox{-1pt}{\rule{6pt}{.3pt}\hspace*
{0pt}\rule{.3pt}{8pt}\hspace*{3pt}}}
\newcommand{\sro}{Schr\"{o}dinger\ }
\newcommand{\bm}{\boldmath}
\newcommand{\vol}{\omega}

\newcommand{\bd}{\mbox{\bf d}}
\newcommand{\bder}{\mbox{\bm $\der$}}
\newcommand{\bI}{\mbox{\bm $I$}}

\newcommand{\be}{\beta} 
\newcommand{\ga}{\gamma} 
\newcommand{\de}{\delta} 
\newcommand{\Ga}{\Gamma} 
\newcommand{\gmu}{\gamma^\mu}
\newcommand{\gnu}{\gamma^\nu}
\newcommand{\ka}{\kappa}
\newcommand{\hka}{\hbar \kappa}
\newcommand{\al}{\alpha}
\newcommand{\lapl}{\bigtriangleup}
\newcommand{\psib}{\overline{\psi}}
\newcommand{\Psib}{\overline{\Psi}}
\newcommand{\derts}{\stackrel{\leftrightarrow}{\der}}
\newcommand{\what}[1]{\widehat{#1}}

\newcommand{\bx}{{\bf x}}
\newcommand{\bk}{{\bf k}}
\newcommand{\bq}{{\bf q}}

\newcommand{\omk}{\omega_{\bf k}}


\section{Introduction }

Contemporary approaches to 
 the 
quantum theory of gravity 
(for a recent review and references see, e.g., \cite{rovelli}) 
either start 
from the classical theory 
trying to quantize it 
  by means of  
various available techniques,  
or  
seek to 
build a more general framework of  which 
 the 
quantum theory of gravity  would be a byproduct or a limiting case. 
The attempts of the first kind include 
quantum geometrodynamics based 
on the Wheeler-De Witt equation,  
Ashtekar's  program of nonpertubative canonical quantum gravity  
and 
approaches based on the path integral.   
 String/M- theory 
and, in a sense, 
models inspired by non-commutative geometry 
represent 
approaches of the  second kind.     
The goal 
of all this is 
a synthesis of quantum theory and general relativity 
 either 
in the sense of 
a fusion into a ``quantum general relativity''  
 or 
incorporation   
 into 
a more general unifying theory of 
all interactions. 

The existing attempts to {\em quantize} gravity 
are known 
to be confronted by 
both 
 the problematic mathematical meaning of the involved 
 constructions 
and 
the 
conceptual questions  
originating in 
difficulties of reconciling 
the fundamental principles of quantum theory with those 
of general relativity 
(see e.g. \cite{isham} for a review 
and further references).\footnote{Note, however, that 
 the 
quantum field theory of gravity can be formulated 
in the low energy domain 
as an effective field theory \cite{donoghue}. }    
In particular, 
a 
 distinct 
role of the time 
dimension 
in 
 the 
probabilistic interpretation of quantum theory 
and in 
 the 
formulation of  
quantum evolution laws 
seems to contradistinguish from 
the equal rights status 
of space-time dimensions in the theory of relativity.     
 Moreover, 
{\em canonical quantization} 
 is 
preceded by the Hamiltonian 
formulation which requires the singling out of 
a 
time parameter or, more 
technically, the global hyperbolicity of 
space-time. The latter 
assumption, however, 
looks rather unnatural 
for 
the 
quantum fluctuating 
``space-time foam''  of quantum gravity,    
 that may indicate 
 that 
 the 
applicability 
 to quantum gravity  
of procedures based on, or inspired by,   
the standard Hamiltonian framework  
 can be rather limited. 


 These difficulties could be overcome 
if one 
 would have in our disposal 
a quantization procedure in field theory 
which  
 does not 
so sensibly depend on 
 the 
 singling out of  
 a  
time parameter.  
However,    
such a ``timeless'' procedure, 
if 
understood as a 
generalized version of canonical quantization,    
clearly  
requires 
a version of 
canonical formalism without 
 the 
 distinct 
role of the time dimension 
and thus  independent of the 
picture of fields 
as infinite dimensional 
systems evolving 
in time from the initial Cauchy data 
given on a space-like hypersurface.

Fortunately, although 
this 
 seems to be 
not yet commonly 
 known 
in theoretical physics,   
the   
Ham\-ilton\-ian-like reformulations  
of the field equations 
which could be appropriate for 
 a 
``timeless'' version of canonical formalism 
 have been known 
in the calculus of variations 
already at least since the thirties. 
In the simplest of them, 
the so-called De Donder-Weyl (DW) formulation 
\cite{dw,dickey},  
the 
 Euler-Lagrange 
field equations 
are written in the form 
\beq
\der_\mu y^a  = \frac{\der H}{\der p^\mu_a}, 
\quad 
\der_\mu  p^\mu_a=- \frac{\der H}{\der y^a } , 
\eeq
where 
$y^a$ denote  field variables, 
$p_a^\mu:=\der L / \der (\der_\mu y^a)$ 
are 
 to be referred to as  
{\em polymomenta},   
$H:=\der_\mu y^a p^\mu_a -L$ is a 
function of $(y^a,p^\mu_a,x^\nu)$ 
to be 
called 
the {\em DW Hamiltonian function},   
and  
$L=L(y^a,\der_\mu y^a,x^\nu)$ 
is a Lagrangian density. 
In this formulation 
the analogue of the configuration space is 
a 
finite dimensional
space of space-time and field variables 
$(x^\nu,y^a)$  
and the analogue of the extended phase space is 
a 
finite dimensional 
space of variables $(y^a, p_a^\mu,x^\nu)$ 
 called 
the (extended) {\em polymomentum phase space}.   
Note, that 
in 
formulation (1.1) fields 
are 
essentially 
described 
as a sort of 
 multi-parameter generalized 
Hamiltonian systems rather than 
as infinite dimensional mechanical 
systems, 
as in 
the 
standard Hamiltonian formalism. 
In doing so 
the DW Hamiltonian function $H$, which 
thus far does not appear to have any evident 
physical interpretation, 
in a sense controls  
 the 
space-time {\em variations}   
of  fields,     
 as 
specified by 
equations (1.1),      
rather than their time {\em evolution}. 
However, the latter is  implicit in (1.1)  in the case 
of hyperbolic  field equations for which 
the Cauchy problem can  
be posed.  

It is interesting to note 
that for the set of DW canonical 
equations (1.1) there exists 
an analogue of the Hamilton-Jacobi 
theory. The corresponding DW Hamilton-Jacobi equation 
  \cite{dw,vonrieth} 
is formulated 
for $n$ ($n$=the number of space-time dimensions) functions 
$S^\mu=S^\mu(y^a,x^\mu)$:   
\beq
\der_\mu S^\mu + H (x^\mu,y^a,p_a^\mu=\der S^\mu/\der y^a)=0 ,    
\eeq 
and naturally leads to the question 
 as to 
which formulation of quantum field theory 
could yield 
this field theoretic 
Hamilton-Jacobi equation 
in the classical limit.


It should be mentioned that the DW 
formulation 
is a particular case of more 
general {\em Lepagean} canonical theories 
 for 
fields 
which 
differ by 
 the 
definitions of polymomenta and  
analogues of Hamilton's canonical function (both essentially follow 
from different choices of 
 the 
Lepagean equivalents of 
the Poincar\'e-Cartan form; for further details and references 
see \cite{dedecker,krupka,gotay-ext}   
and reviews quoted in \cite{dw}).  
All 
theories of this type 
 treat space and time variables on equal footing and 
are finite dimensional in the sense that the 
corresponding analogues 
 of the configuration 
and the phase space are finite 
dimensional.  In addition, they all reduce to the Hamiltonian 
formalism of mechanics when $n=1$. 
In a sense, all these formulations are 
intermediate between the Lagrangian formulation and 
the canonical Hamiltonian one. They still keep space-time 
variables indistinguishable but already possess essential 
features of the Hamiltonian-like description such as  
a reference to the first order form of the field equations 
and a Legendre transform.   
Moreover, there seem to be  intimate relations, 
not fully studied as yet, between 
structures of the canonical Hamiltonian formalism 
and those of the  
Lepagean formulations. 
For this reason,  let us  
refer to these formulations 
and 
 the 
related structures as ``{\em pre-canonical\,}''. 
Further justification of the term will be given in 
Conclusion.



\newcommand{\oldtexta}{  
For the reasons 
 discussed 
in Conclusion 
 let us 
refer to 
 approaches 
based on 
formulations 
of the field equations 
 similar to (1.1) as ``pre-canonical''.  
The main reason is that these 
 approaches 
 are in a sense intermediate between 
the 
Lagrangian 
 one 
and the standard 
canonical 
Hamiltonian 
one. 
 } 


It is worthy of noticing that pre-canonical formulations 
typically have different regularity conditions than the standard 
Hamiltonian formalism. For example, the DW formulation (1.1) implies 
that $\det || \der^2 L / \der_\mu y^a \der_\nu y^b || \neq 0$ 
which is different from the usual requirement that 
$\det || \der^2 L / \der_t y^a \der_t y^b || \neq 0$.  
As a result, the constraints understood as obstacles to the corresponding 
generalized Legendre transforms  
 $\der_\mu y^a \rightarrow p^\mu_a$ 
have 
 a 
quite different structure 
so that  singular theories from the point of view of the conventional 
formalism can be regular from the pre-canonical point of view 
(as, e.g., the Nambu-Goto string \cite{romp98}) or vice verse 
(as, e.g., the Dirac spinor field \cite{vonrieth}).  
This opens an yet unexplored 
possibility 
of avoiding the constraints analysis when quantizing 
within the pre-canonical framework 
by choosing for a given theory an appropriate non-singular 
Lepagean Legendre transform (in fact, this possibility is 
exploited below when quantizing General Relativity without 
any mention of constraints).

The idea of quantization proceeding from the  DW canonical theory 
for fields dates back to Born and Weyl 
\cite{bw} 
but has 
not received much attention 
since then 
(see, however, \cite{attempts}). 
Obviously,  quantization  
 needs 
more than just 
 an existence of 
a Hamiltonian-like formulation 
of the field equations.  Additional structures,   
such as the Poisson bracket (for  canonical or  deformation 
quantization),  the symplectic structure 
(for  geometric quantization),   
and 
a 
Poisson bracket formulation of the 
field equations (in order to 
formulate 
the 
quantum dynamics law) are 
necessary. 
In spite of a number of earlier 
attempts \cite{oldbrackets} 
and 
related 
developments in the geometry of classical field 
theory based on the 
(Hamilton-)\-Poincar\'e-Cartan, 
 or 
 multisymplectic, form 
\cite{sternberg,crampin,gimmsy} 
(see \cite{gimmsy} for more extensive references)   
 and  
 on 
 another 
generalization of the symplectic structure 
due to G\"unther \cite{guenther},    
a construction which could be suitable 
 as a starting point of  quantization   
was found  only recently   
\cite{bial94,romp98,bial96}. 
It leads to graded Poisson brackets 
defined on differential forms 
which 
represent  in this case 
dynamical variables.    
 From a 
 more mathematical perspective,   
various  relevant 
 jet-bundle-theoretic constructions  
and 
generalizations of the symplectic geometry 
have been 
 studied 
recently 
 by a number of authors 
 \cite{
echev1,sardan, 
deleon,marsden98,hrabak,cantrijn,norris}   
who have significantly improved our knowledge of the 
underlying geometric structures of classical field 
theories,  
yet to be employed by physicists.   

The elements of 
quantization  
in field theory 
 based on the 
aforementioned brackets 
 on differential forms 
have been 
discussed in  
\cite{bial94,qs96,bial97,lodz98} 
and will be briefly  
summarized 
below.    
It should be noted that 
 so far 
this is 
 rather a preliminary 
 approach   
some fundamental aspects of which, 
 including  a proper interpretation and 
a connection with the standard formalism of quantum field theory,  
are yet to be clarified. 
However,  
 the approach 
possesses 
 a distinct 
esthetic  attraction, 
intriguing 
 features,     
and as yet unexplored potential 
which 
make it %
worthy of further endeavors. 
The intrinsic finite dimensionality 
and 
manifest covariance  
 could  
make  
 the 
pre-canonical framework,  
if successful, 
 a suitable  
complement 
to the presently available concepts 
and techniques of quantum field theory.  

The main purpose of the present paper 
is to  discuss 
a preliminary application 
of 
 pre-canonical framework 
to the problem 
of quantization of General Relativity.  
We first, in Section 2, summarize 
basic aspects of 
 pre-canonical formalism 
and  
quantization 
based on the DW theory 
and then, in Section 3, 
apply  
it 
to General Relativity. 
Discussion 
and 
concluding remarks 
are presented in Section 4.

\section{Pre-canonical formalism and quantization 
based on 
DW theory }

In this section we briefly summarize  basic  elements 
of the analogue of  canonical formalism based on 
DW theory 
and then 
 outline 
elements of 
quantization based on this formalism.

\subsection{Classical theory }  
The mathematical structures underlying 
 the DW form of the field equations, 
 which can be suitable for quantization, 
 have been 
studied in our previous papers 
\cite{romp98,bial96} to which we refer for more details.

The 
analogue of the  
Poisson bracket for the DW formulation 
can be deduced from the object 
called the {\em polysymplectic form }  
which in local coordinates 
can be written in the form\footnote{Note, that this object can be understood 
as the equivalence class modulo  forms of the highest horizontal 
degree $n$, see \cite{romp98} for more details. 
Henceforth we 
 denote  
$ 
\omega := dx^1\we ... \we dx^n , \quad 
\omega_\mu := \der_\mu\inn \omega = 
(-1)^{\mu -1}dx^1 \we ... \what{dx^\mu} ... \we dx^n  
$, 
$dx\uind{\mu}{p} :=  
dx ^{\mu_1}\we ...\we dx^{\mu_p} $,  
and $\{z^M \} :=  \{y^a,p_a^\mu,x^\mu \} $.} 
 \[ 
\Omega = - dy^a \we dp_a^\mu \we \omega_\mu 
 \] 
and is viewed as a field theoretic generalization
 of the symplectic form 
within the DW formulation. 
 Note that if  
$\Sigma$,  
$\Sigma\!:\!(y\!=\!y_{in}(\bx), \, t\!=\!t_{in})$, 
denotes the Cauchy data surface 
in the covariant configuration space 
the standard symplectic form in field theory, $\omega_S$,  
 can be expressed as 
 the integral  of  the pull-back 
of $\Omega$ to $\Sigma$, $\Omega|_\Sigma$, 
 i.e.   
$\omega_S = \int_\Sigma \Omega|_\Sigma$.    
 The form $\Omega$ allows us to set up 
relations between 
$p$-forms $\ff{p}$ 
and  $(n-p)$-multivectors 
(or more general algebraic operators 
 of degree $-(n-p)$ 
on the exterior algebra)  
 $\xx{n-p}$: 
\beq 
\xx{n-p} \inn \Omega 
= d \ff{p} . 
\eeq   
 Then the graded Poisson brackets  
of  
horizontal 
forms 
$ 
\ff{p}: 
= \frac{1}{p!}  
F\lind{\mu}{p}(z^M)
 dx\uind{\mu}{p} 
$, { }  
($p=0,1,...,n$), 
can be defined 
by 
\beq
\pbr{\ff{p}{}_1}{\ff{q}{}_2} := (-)^{n-p} \xx{n-p}{}_1 \inn d \ff{q}{}_2. 
\eeq
     Hence 
the bracket of a $p$-form with a $q$-form is a form of degree 
$(p+q-n+1)$, where $n$ is the dimension of the space-time. 
 Consequently, the subspace of forms of degree $(n-1)$  
is closed with respect to the bracket, as well as the subspace of 
forms of degree $0$ and $(n-1)$. 

The construction 
leads to 
a 
hierarchy of algebraic structures which are graded 
generalizations of 
 the 
Poisson algebra in mechanics 
\cite{romp98,bial96}. 
Specifically, 
on a small subspace of horizontal forms called Hamiltonian forms 
(i.e. those which can be associated   
by  relation (2.1)      
with multivectors)   
we obtain the structure 
of a Gerstenhaber algebra. 
Recall, that the latter is a triple 
${\cal G}=({{\cal A},\pbr{\ }{\,}, \cdot})$, 
where ${\cal A}$ is a graded commutative associative 
algebra with the product operation $\cdot$, $\pbr{\ }{\, }$ is a 
graded Lie bracket 
fulfilling the graded Leibniz rule with respect to the product $\cdot$, with 
the degree of an element $a$ of ${{\cal A}}$ with respect to the bracket 
operation, $bdeg(a)$, 
and the degree of $a$ with respect to the 
product $\cdot$, $pdeg(a)$, related as $bdeg(a)=pdeg(a)+1$.  
In our case the bracket operation is a graded Lie bracket closely related 
to the Schouten-Nijenhuis bracket of multivector fields (the latter is 
related to our bracket in the similar  
way as the Lie bracket of vector fields is related to the Poisson bracket).  
Correspondingly,  the graded commutative multiplication 
 $\bullet$  
(the ``co-exterior product''),   
with respect to which the space of Hamiltonian 
forms is stable,  
is given by  
 \[ 
F\bullet G:=*^{-1}(*F\we*G), 
\] 
where $*$ is the Hodge 
duality operator. 
On more general (``non-Hamiltonian'') forms 
a 
non-commuta\-tive 
(in the sense of Loday's ``Leibniz algebras''  
\cite{loday})    
higher-order (in the sense of a higher-order 
analogue of the graded Leibniz rule 
replacing the standard Leibniz rule 
in the 
definition)  
generalization of a Gerstenhaber algebra 
appears 
\cite{bial96}.  

The bracket defined in (2.2) 
enables  
 us 
to identify 
 pairs of 
canonically conjugate variables 
and to represent the DW canonical equations 
in (generalized) Poisson bracket formulation.  
Namely, the appropriate notion of 
canonically conjugate variables in the present 
context is suggested by considering brackets 
of horizontal forms 
of the kind 
$y^a dx^{\mu_1}\we ... \we dx^{\mu_p}$ 
and $p_a^\mu \der_\mu \inn\der_{\mu_1}\inn ... 
\der_{\mu_q}\inn \omega $  
with $(p-q)\geq 0$.  
In particular, 
in the Lie subalgebra of 
 Hamiltonian forms of degree 
$0$ and $(n-1)$ the non-vanishing 
 ``pre-canonical'' 
brackets 
take the form \cite{romp98}
\beqa 
\pbr{p_a^\mu\omega_\mu}{y^b}
&=& 
\delta^b_a ,  
 \nn \\
\pbr{p_a^\mu\omega_\mu}{y^b\omega_\nu}
 &=&
\delta^b_a\omega_\nu,  
 \\ 
\pbr{p_a^\mu}{y^b\omega_\nu}
 &=&
\delta^b_a\delta^\mu_\nu    
\nn 
\eeqa 
and 
are seen to 
reduce to the canonical Poisson bracket in mechanics 
when $n=1$. 
Hence, the 
 pairs of variables entering  the brackets (2.3) 
can be viewed as canonically conjugate 
 with respect to the graded Poisson 
bracket (2.2). 
Note that no dependence on $x^\mu$-s is implied in 
(2.3) so that the above brackets 
should be viewed rather as  
``equal-point'', as opposite to the conventional 
``equal-time'' ones.  

Now, by considering the brackets of 
canonical variables 
with $H$ or $H\omega$ 
one can write the DW canonical field equations 
(1.1) in graded Poisson bracket formulation, e.g.  
 \beqa 
\bd(y^a\omega_\mu) 
&=&  \pbr{H\omega}{y^a\omega_\mu}, 
\nn \\ 
\bd(p^\mu_a \omega_\mu) 
&=& \pbr{H\omega}{p^\mu_a \omega_\mu},  
\eeqa 
where $\bd$ is the total exterior differential 
such that, for example, 
$\bd y = \der_\mu y(x) dx^\mu $.  
The 
DW canonical equations written in the form (2.4) 
point to the fact that 
 the 
 type of the space-time variations  
controlled 
by $H$ 
is 
intimately related to the exterior 
differentiation. This generalizes to the 
present formulation of field theory the 
familiar statement 
that  
Hamiltonian generates 
 a 
time evolution. Note that the form (2.4) 
of 
the 
canonical 
field equations 
underlies our hypothesis 
(2.6)  
 regarding 
 the form of 
 a 
 generalized Schr\"odinger equation 
 within 
the 
pre-canonical approach (see \cite{bial94,qs96,bial97}).

\subsection{Quantization }   
Quantization of a Gerstenhaber algebra 
${\cal G}$ or its above-mentioned generalizations 
 would be a difficult mathematical problem. 
We may even  need 
to modify the notion of quantization or deformation to treat this 
problem properly \cite{flato}.   
This is due to the fact that $bdeg(a)\neq pdeg(a)$ for $a \in {\cal G}$. 
An attempt to adopt geometric quantization to the present case 
also faces this problem already on the  pre-quantization level. 
Fortunately, in physics we usually do not need to quantize the whole  
Poisson algebra. It is even impossible in the sense of 
canonical 
quantization, as it follows from the 
Groenewold-van Hove ``no-go'' theorem 
\cite{emch,gotay-q}.     
In fact,  quantization of 
a 
small Heisenberg subalgebra of the canonical 
brackets often suffices.  
All we need to know about the rest of a Poisson algebra is essentially 
that ``it is there''. 

Thus, as 
the 
first step 
it seems reasonable 
 to   restrict our attention to a small subalgebra of graded Poisson brackets 
which 
is 
similar to the Heisenberg subalgebra of a Poisson algebra 
in mechanics.  A natural candidate is the 
subalgebra of canonical brackets in the 
Lie subalgebra of Hamiltonian forms of degree $0$ and $(n-1)$, 
see eqs. (2.3).  
The scheme of 
field quantization discussed in \cite{bial94,qs96,bial97}  
is essentially based on  quantization of this 
small subalgebra by the Dirac correspondence rule. 
The following realization of operators corresponding 
to the  quantities involved in (2.3)   
 was proposed 
\beqa
\widehat{p_a^\mu\omega_\mu}&=& i \hbar 
 \,\der / \der y^a 
 , \nn \\
\hat{p}{}_a^\nu &=&  - i \hbar  \kappa \ga^\nu \
  \der / \der y^a 
  ,  \\ 
\widehat{ \omega}_\nu&=&  - \kappa^{-1} \ga_\nu 
  ,  \nn 
\eeqa 
where $\gamma^\mu$ are 
the 
imaginary units of   
the Clifford algebra 
of the space-time manifold and    
the parameter $\kappa$ of the dimension 
{(length)}$^{-(n-1)}$ is 
required by 
the dimensional consistency of (2.5). 
A possible  
identification 
 of $\kappa$ with   
the ultra-violet cutoff 
or 
the fundamental length scale 
 quantity was discussed in \cite{qs96,lodz98}. 
Note, that 
realization (2.5) is essentially inspired  by 
the relation between the 
 Clifford algebra and 
the 
endomorphisms of the exterior algebra.     
 A crucial assumption 
underlying the proof 
 that operators in (2.5) fulfill 
 the 
commutators following from (2.3) 
is  that the composition law of operators 
 implies  
 the symmetrized product of $\gamma$-matrices. 

The realization (2.5) suggests that 
quantization 
of DW formulation, 
viewed as 
a multi-parameter generalization 
of the standard  Hamiltonian formulation with 
a 
single time parameter, 
 results in 
a version, 
or a generalization,   of 
 the 
quantum theoretic formalism in which the hypercomplex (Clifford) 
algebra of the underlying space-time manifold 
 generalizes 
the algebra of complex numbers (=the Clifford algebra of 
 (0+1)-dimensional ``space-time'') 
in quantum mechanics, 
and 
 in which 
$n$ space-time variables 
 being 
treated 
on 
equal footing 
generalize 
the single time parameter. 
In doing so quantum mechanics appears as a special 
case corresponding to $n=1$. 
This 
 philosophy 
 suggests the following 
 generalization of  
the Schr\"odinger equation 
 to the present 
 framework \cite{qs96,bial97,lodz98}  
\beq
\label{seqcl}
i \hbar \kappa \gamma^\mu \der_\mu \Psi = \what{H} \Psi, 
\eeq
where $\widehat{H}$ is the operator corresponding to the 
DW Hamiltonian function, the constant $\kappa$ 
of the dimension ({length})$^{-(n-1)}$ appears again on dimensional 
grounds, and $\Psi=\Psi(y^a,x^\mu)$ is 
the 
wave function over 
the 
covariant 
configuration space of 
 field and space-time variables. 

The generalized Schr\"odinger equation (2.6)  
satisfies several aspects of the correspondence 
principle   
\cite{bial97,lodz98}. In particular, it leads, at least 
in the simplest case of scalar fields, 
to the DW canonical equations (1.1) for the mean values 
of appropriate operators (the Ehrenfest theorem) 
and reduces to 
the DW Hamilton-Jacobi equation (1.2) 
(with some additional conditions) 
in the classical limit.    
For an application of the 
presented scheme 
to the case of scalar fields 
see \cite{bial97,lodz98}.  


\section{Quantizing General Relativity }

In this section we first outline curved space-time generalization 
of 
 the scheme 
presented in  sect. 2.2 and then discuss  its further 
 application to quantization of General Relativity. The latter 
requires the DW Hamiltonian formulation of General Relativity which 
is discussed in sect. 3.2. 
This framework leads 
to a covariant 
Schr\"odinger equation which is supposed to describe 
quantum  General Relativity and a supplementary condition which 
introduces a background geometry 
in a self-consistent with the underlying quantum 
dynamics way.  

\subsection{ Curved space-time generalization } 
To apply the above 
framework 
to General Relativity we 
first need to extend 
it 
to curved space-time with  metric 
$g_{\mu\nu}(x)$. 
The extension of the generalized Schr\"odinger 
equation (2.6) to 
curved space-time is  similar to that of the Dirac equation, i.e. 
\beq  
i\hbar\kappa 
\gamma^\mu (x) \nabla_\mu \Psi 
= \what{H} \Psi 
\eeq
where 
$\what{H} $ is 
an 
operator form of 
the DW Hamiltonian function    
and 
$\nabla_\mu$ is 
 the 
covariant derivative, 
$\nabla_\mu :=  \der_\mu + \theta_\mu (x)$.   
We introduced 
$x$-dependent $\gamma$-matrices 
which fulfill 
\beq
\gamma_\mu(x) \gamma_\nu(x) + \gamma_\nu(x) \gamma_\mu(x)
=2 g_{\mu\nu}(x)     
\eeq 
and can be expressed with the aid  of 
vielbein fields 
$e_\mu^A (x)$,  such that 
\beq
g_{\mu\nu}(x) = 
e_\mu^A (x) e_\nu^B (x) \eta_{AB}, 
\eeq 
and the 
Minkowski space Dirac matrices $\gamma^A$, 
$\gamma^A\gamma^B + \gamma^B\gamma^A 
:= 2 \eta^{AB}$,    
\[
\gamma^\mu (x):= e^\mu_A (x) \gamma^A. 
\]

 If $\Psi$ is a spinor 
then 
$\nabla_\mu =  \der_\mu + \theta_\mu$ 
 is the spinor 
covariant derivative
with 
 \[ 
\theta_\mu = \frac{1}{4} \theta_{AB}{}_\mu \gamma^{AB},
\]  
$\gamma^{AB}:= \half (\gamma^A\gamma^B - \gamma^B\gamma^A)$, 
 denoting the spinor connection whose components 
are known to be given by 
\beq
\theta^A{}_{B\mu} = 
 e^A_\al e^\nu_B \Ga^\al{}_{\mu\nu} 
 -e^\nu_B \der_\mu e^A_\nu .
\eeq

For example, interacting scalar fields $\phi^a$ on 
a curved background 
are described by 
the Lagrangian density  
 \[
{\cal L}=  \frac{1}{2}\sqrt{g} \{ \der_\mu \phi^a \der^\mu \phi_a 
 - U(\phi^a) 
- \xi R \phi^2 \} ,  
\]  
where $g:={|\rm det}(g_{\mu\nu})|$.  
 It gives rise to the following  
 expressions of polymomenta and the DW Hamiltonian density   
\[ 
p^\mu_a := \frac{\der {\cal L}}{\der(\der_\mu \phi^a)} =  
\sqrt{g}\der^\mu \phi_a ,  
\quad 
\sqrt{g} H  = 
 \frac{1}{2\sqrt{g}}p^\mu_a p^a_\mu + 
\frac{1}{2}\sqrt{g} \{ U(\phi)
+ \xi R \phi^2  \} 
\]
 for which  the corresponding operators can be found to take the form 
 \beqa 
\what{p}{}^\mu_a &=&  
- i\hbar\kappa \sqrt{g} \gamma^{\mu}\frac{\der}{\der \phi^a} , 
  \nn \\ 
\what{{H}} 
&=& -\frac{\hbar^2\kappa^2}{2}
\frac{\der^2}{\der \phi^a \der \phi_a} 
+  \frac{1}{2}
\{ 
U(\phi) 
+ \xi R \phi^2 
\}  
. 
 \eeqa

\subsection{Pre-canonical approach to quantum General Relativity }  

In the context of General Relativity viewed as a field theory  
the 
metric 
$g_{\al\beta}$ 
(or 
the 
vielbein $e_A^\mu$) 
 is the field variable. 
Hence, 
 according to the pre-canonical scheme,   
the wave function is  
a function 
of space-time and metric (or vielbein) variables, i.e. 
$\Psi= \Psi (x^\mu, g^{\al\beta})$ (or $\Psi= \Psi (x^\mu,e_A^\mu)$).    
To 
formulate an analogue 
of the Schr\"odinger equation for this wave function we  need 
$\gamma$-matrices which fulfill  
\beq
\gamma^\mu \gamma^\nu + \gamma^\nu \gamma^\mu
= 2 g^{\mu\nu}    
\eeq
 and 
can be related 
to the 
 Minkowski
space $\gamma$-matrices 
$\gamma^A$  
by means of 
the 
vielbein components:
$\gamma^\mu := e_A^\mu \gamma^A$,  
where 
 $g^{\mu\nu}=: e_A^\mu e_B^\nu\eta^{AB}$. 
Note that,  as opposite to the theory on curved 
background,  no explicit 
dependence on 
space-time variables is present 
in the above formulas; 
instead, $e_A^\mu $, $\gamma^\mu$ and $g^{\mu\nu}$  
are viewed as  
  the 
fibre coordinates in the corresponding bundles 
over the space-time. 

Now,  modelled after (3.1),  
the following (symbolic form of the) 
generalized Schr\"o\-din\-ger equation 
for the wave function of quantized gravity can be 
put forward 
\beq
i\hbar\kappa 
\what{ \mbox{$\hspace*{0.0em}e\hspace*{-0.3em}
\not\mbox{\hspace*{-0.2em}$\nabla$}$}} 
 \Psi = 
\what{{\cal H}\hspace*{-0.0em}} \Psi  ,   
\eeq 
where   
$\what{{\cal H}}$ is the operator form 
of the DW Hamiltonian 
 density  
of gravity, $\what{{\cal H}}:= \what{e H}$,    
an explicit form of which remains  to be constructed, 
$e:= |{\rm det}(e^A_\mu)|, $ 
and 
$\what{\mbox{$\hspace*{0.0em}\not\mbox{\hspace*{-0.2em}$\nabla$}$}}$  
denotes the quantized Dirac operator in the sense that 
the corresponding connection coefficients are 
replaced by appropriate differential operators  
(c.f., e.g.,  eqs. (3.15), (3.16) and (3.21) below).   
Note also, that 
in the context of quantum gravity it 
seems 
 to be 
very natural to identify the parameter $\kappa$ in (3.7) 
with the Planck scale quantity, i.e. 
$\kappa \sim l_{\mbox{\footnotesize Planck}}^{-(n-1)}$.  


If the wave function in (3.7) is  spinor  
then the covariant derivative operator   
$\what{\nabla}{}_\mu $ contains the spinor connection which 
on the classical level involves 
 the term with  
space-time derivatives 
of vielbeins  (c.f. eq. (3.4)) 
which cannot be expressed in terms of 
quantities 
of the metric formulation. 
Consequently, the spinor nature of  equation (3.7)  
necessitates the use of the vielbein formulation of General 
Relativity.
However,  
 since no  suitable DW formulation of General Relativity 
 in  vielbein variables 
is available  
so far,   
the subsequent 
consideration  
 will be based on 
the metric formulation   
   while 
the analysis based on the vielbein 
formulation is postponed to a future publication.

\subsubsection{DW formulation of the Einstein equations }  
A 
suitable DW-like   
formulation of 
General Relativity in metric variables 
was presented earlier by Ho\v rava \cite{horava}. 
 In this formulation 
field variables are chosen to be 
the metric density components 
$ h^{\alpha\beta}:= \sqrt{g}g^{\al\be}$ 
and the role of polymomenta is 
taken on by  the following combination of 
the Christoffel symbols 
\beq 
Q^\al_{\beta\gamma} := 
 \frac{1}{8\pi G}( \delta^\alpha_{(\beta}\Gamma^\delta_{\gamma)\delta} 
  - \Gamma^\al_{\beta\gamma}) .   
\eeq 
Respectively, 
the DW Hamiltonian 
{ density}  ${\cal H}:= \sqrt{g} H$   
 assumes 
 the form 
\beq
{\cal H} 
(h^{\alpha\be}, Q^\al_{\beta\gamma}) :=  
 8\pi G\, 
h^{\alpha\ga} \left ( 
Q^\de_{\al\be } Q^\be_{\ga\de }+ 
\frac{1}{1-n}\, Q^\be_{\al\be }Q^\de_{\ga\de } \right )
+(n-2) \Lambda \sqrt{g}  
\eeq  
which is 
essentially the 
truncated Lagrangian 
density of General Relativity 
 written in terms 
of variables $h^{\alpha\be}$ and 
$Q^\al_{\beta\gamma}$. 

Using these variables the Einstein field 
equations are formulated in 
DW Hamil\-ton\-ian form 
as follows
\beqa    
\der_\al h^{\be\ga}
&=& 
\der {\cal H}  / \der Q^\al_{\be\ga} , \\ 
\der_\al Q^\al_{\be\ga}
&=& 
- \der {\cal H}   / \der h^{\be\ga}  ,  
\eeqa  
where eq.~(3.10) is equivalent  to the well-known expression 
of the Christoffel symbols in terms of the metric 
while 
eq.~(3.11)  
 yields 
the vacuum  
Einstein equations 
in terms of  the Christoffel symbols.

\subsubsection{Naive pre-canonical quantization }  
Now, to quantize 
General Relativity in 
 DW formulation we 
can 
{\em formally }  
follow   
the curved space-time version of our scheme,  
sect. 3.1.  
This leads to the 
 following operator form 
of polymomenta $Q^\al_{\be\ga}$ 
\beq
\what{Q}{}^\al_{\be\ga} = -i\hbar \kappa 
\gamma^\alpha 
\left \{  \sqrt{g} 
\frac{\der}{\der h^{\beta\gamma}} \right \}_{ord} 
\eeq 
up to 
 an 
 ordering ambiguity of the expression inside 
 the curly brackets.  
By substituting this expression to (3.9) 
and performing some 
manipulations 
using the assumption of 
the ``standard'' ordering 
(that differential operators 
are all collected to the  
right) 
and  
relation  (3.6) for curved 
$\gamma$-matrices 
we obtain the following 
operator form of the 
DW Hamiltonian 
density 
\beq
\what{{\cal H}} 
= - 8\pi G\, 
\hbar^2\kappa^2 \frac{n-2}{n-1}  
\left \{  
 \sqrt{g} 
h^{\al\ga}h^{\be\de}\frac{\der}{\der h^{\al\be}} 
\frac{\der}{\der h^{\ga\de}} \right \}_{ord} 
+ (n-2)  \Lambda 
 \sqrt{g} 
\eeq  

However, 
it should be pointed out that the above procedure 
is rather of heuristic nature and  
  requires a mathematical justification. 
Namely, according to (3.8) classical polymomenta 
$Q^\al_{\beta\gamma}$ transform as 
connection coefficients 
while the operator we associated with them in (3.12)  
is a tensor.   
Moreover, 
the classical DW Hamiltonian density in (3.9) 
does not transform as a scalar density,  
while the operator constructed in (3.13)  
is a scalar 
density.   
 It is thus natural to 
inquire 
whether, 
or in which sense, 
the whole procedure 
is meaningful. 

\subsubsection{Covariant Schr\"odinger equation 
 for quantized gravity 
and the ``bootstrap condition'' }
The answer 
  to the above question 
refers to   the 
 observation 
that 
 canonical 
quantization procedures are 
 in fact %
generally   
performed with respect to a specific 
reference frame 
and require  
``covariantization'' 
as a subsequent step. 
Note now,  
that the consistency of 
expressions (3.12) and (3.13)  
with classical transformation laws 
could be achieved by 
adding an 
auxiliary  
term in (3.12) which transforms as a connection 
	and can be interpreted as 
appearing 
due to 
 an auxiliary  
connection corresponding  to a chosen 
coordinate system or a background. 
Then, using the expression of 
the 
Christoffel symbols in terms of the polymomenta 
$Q^\al_{\be\ga}$ (c.f. (3.8)) 
\beq
\Gamma^\al_{\be\ga}= 8\pi G 
\left ( 
\frac{2}{n-1} 
\delta^\al_{(\beta } Q^\delta_{\ga)\delta} 
- Q^\al_{\be\ga} 
\right ) 
\eeq 
we are led  to the following 
 ordering dependent 
operator form of the Christoffel symbols 
\beq  
\what{\Gamma}{}^\al_{\be\ga} = 
- 8\pi i G\hbar\kappa \sqrt{g} 
\left 
(\frac{2}{n-1} \delta^\al_{(\beta } \gamma^\sigma 
\frac{\der}{\der h^{\ga ) \sigma }} 
- \ga^\al \frac{\der}{\der h^{\beta \ga}} 
\right ) 
+ \tilde{\Gamma}{}^\al_{\be\ga}(x) 
\eeq 
where the auxiliary connection 
is denoted 
as $\tilde{\Gamma}{}^\al_{\be\ga}(x)$. 
However, if we keep 
the arbitrary term 
$\tilde{\Gamma}{}^\al_{\be\ga}(x)$ 
it  
will enter 
the 
final results thus  
leading to a rather unappealing 
background dependent theory.  

On the other hand, we can 
notice that our 
operators 
essentially arise from  
the ``equal-point'' commutation relations  
 (c.f. eqs.~(2.3))  
and thus can be viewed as 
 locally 
defined  ``in a point''.  
Now, 
in an infinitesimal vicinity 
of a point 
we always can chose 
a local coordinate system 
in which the auxiliary  connection 
$\tilde{\Gamma}{}^\al_{\be\ga} (x) $ 
vanishes and then 
think of this system as the one 
which is 
 actually 
meant  when 
writing 
expression (3.12) for operators $\what{Q}$.  
This, however,  
 when consistently implemented,  
requires  
a subsequent  
"patching together" procedure which 
is likely to  
lead to extra terms in our 
generalized Schr\"odinger 
equation 
due to %
the 
connection 
involved in "patching together". 

To cope with the above-mentioned problems 
we 
proceed as follows. Let us first 
write the generalized Schr\"odinger equation (3.7) 
in the local 
coordinate system in the 
vicinity of a point $x$, 
in which $\tilde{\Gamma}^\al_{\be\ga} |_x$ = 0,  
and then  
covariantize it in the simplest way. 
 The first step leads to 
the locally valid equation 
\beq
i\hbar\kappa \sqrt{g}\gamma^\mu 
(\der_\mu + \hat{\theta}_\mu)  
\Psi = 
\what{{\cal H}} \Psi . 
\eeq
Next, the operator 
 form of the spinor 
connection coefficients, 
as follows from 
(3.4) and (3.15),   
is  given by 
\beqa 
\what{\theta}^A{}_{B\mu} &=&  
-8\pi i G \hbar \kappa \sqrt{g} 
e^A_\al e^\nu_B 
\left ( 
\frac{2}{n-1} \delta^\al_{(\mu } 
\gamma^\sigma 
\frac{\der}{\der h^{\nu)\sigma }} 
- \ga^\al \frac{\der}{\der h^{\mu \nu}} 
\right ) 
+\tilde{\theta}^A{}_{B\mu}(x) \nn \\
&=:& ({\theta}^A{}_{B\mu})^{op} + 
\tilde{\theta}^A{}_{B\mu}(x) 
\eeqa  
where $({\theta}^A{}_{B\mu})^{op}$ 
is  a notation introduced for 
the first 
 ordering dependent 
operatorial term. 
Note that in general 
$\tilde{\theta}^A{}_{B\mu}|_x \neq 0 $  
even though 
$\tilde{\Gamma}^\al_{\be\ga}|_x =0$.

Now, to formulate a covariant version of 
(3.16) we notice that 
vielbeins do not 
enter the present DW Hamiltonian 
formulation of General Relativity 
and, therefore, 
within the present consideration 
may (and can only) be 
treated 
as non-quantized classical 
$x$-dependent  
 quantities 
describing a  
reference vielbein field 
which 
accounts 
for  a choice of 
coordinates or a background.   

On another hand, 
the  
bilinear 
combination  of vielbeins 
$e_A^\mu e_B^\nu\eta^{AB}$ is 
the metric 
 $g^{\mu\nu}$ 
which {\em is} a canonical  
variable  quantized 
as an $x$-independent quantity 
(in the ``ultra-Schr\"odinger'' 
picture used here,  
in which all 
space-time 
dependence is 
converted 
%
%
to the wave function 
 while 
the operators are space-time independent).  
Both aspects can be 
 reconciled 
by 
 requiring 
that the bilinear combination of vielbeins 
should be consistent with the 
mean  value  
of the metric, 
i.e. 
\beq  
\tilde{e}{}^\mu_A(x)
\tilde{e}{}^\nu_B(x) \eta^{AB} = 
\left < g^{\mu\nu}\right >(x) ,    
\eeq   
where 
the latter 
is given by 
averaging  over the space of 
 the 
metric components by means of 
the wave function $\Psi(g^{\mu\nu},x^\mu)$:   
\beq
\left <g^{\mu\nu}\right >(x) = 
\int [d g^{\al\be}] 
~\Psib (g,x) g^{\mu\nu} \Psi(g,x)  .  
\eeq 
The 
invariant integration measure in (3.19) 
 can be found to be 
(c.f. \cite{misner}) 
\beq
[d g^{\al\be}] = 
\sqrt{g}^{\,(n+1)} d^{\frac{1}{2} n(n+1)} \!g^{\al\be} . 
\eeq
Thus,   
the background metric 
geometry explicitly appears as 
 a 
result of  
quantum averaging of 
the metric operator $g^{\mu\nu}$,     
while 
the local orientation of 
vielbeins  is 
supposed 
to be exclusively due to a choice of 
 a 
reference 
vielbein field 
(local reference frames)   
on the macroscopic level, 
 which is 
restricted only by the consistency with the 
averaged metric 
 according to 
the ``bootstrap condition'' (3.18).   

Now, 
the covariantized 
version of (3.16)   
can be written 
in the form 
\beq  
i\hbar\kappa 
\tilde{e} 
\tilde{e}^\mu_A(x)\ga^A(\der_\mu + 
\tilde{\theta}_\mu (x)) \Psi  
+ 		
i\hbar\kappa (\sqrt{g} \gamma^\mu {\theta_\mu})^{op} \Psi 
=  \what{{\cal H}}\Psi . 
\eeq 
The 
explicit form of the term 
$(\sqrt{g} \gamma^\mu {\theta_\mu})^{op} $   
can be derived from (3.17) by assuming  the ``standard'' ordering  
in the intermediate calculations 
and replacing the 
 appearing 
therewith 
bilinear combinations   
of vielbeins 
with the metric. 
This procedure yields 
the result 
\beq
(\sqrt{g} \gamma^\mu {\theta_\mu})^{op} 
= - n 
\pi i G \hbar\kappa 
\left \{ \sqrt{g} h^{\mu\nu} 
\frac{\der}{\der  h^{\mu\nu}} \right \}_{ord} 
\eeq 
which is also ordering dependent. 
The $x$-dependent 
reference spinor connection 
term $\tilde{\theta}_\mu (x)$ 
in (3.21) 
is calculated using 
the reference vielbein field 
$\tilde{e}{}^A_\mu (x)$ consistent with (3.18)  
and  the 
classical expression 
\beq
\tilde{\theta}{}_\mu^{AB} (x) =  
\tilde{e}{}^{\al [A} 
\left 
(\half \der_{[\mu} \tilde{e}{}^{B ]}_{\al ]} 
+ \tilde{e}{}^{ B] \be} \tilde{e}{}^C_\mu \der_\be 
\tilde{e}{}_{C \al}  
\right )
\eeq
which is equivalent to (3.4). 
Lastly, 
to distinguish a physically relevant information 
in (3.21)  we need to impose a gauge-type condition on $\Psi$ 
which, in the case of the De Donder-Fock harmonic gauge, 
can be written in the form 
\beq 
\der_\mu \left <\sqrt{g}g^{\mu\nu}\right >(x) = 0. 
\eeq

Thus, we 
conclude 
that within 
the 
quantization based on DW formulation 
the 
quantized gravity 
may be 
described by 
 the 
generalized Schr\"odinger 
equation (3.21), with operators $\what{{\cal H}}$ and 
$(\sqrt{g} \gamma^\mu {\theta_\mu})^{op} $ given 
respectively by 
 (3.13) and  (3.22),  
and 
the 
supplementary 
``bootstrap condition'' (3.18). 
These equations in principle allow us to obtain 
the wave function $\Psi(g^{\mu\nu}, x^\al)$ 
which 
then may 
be interpreted as the probability 
amplitude 
 to find 
the values of the components 
of the metric in the interval 
$[g^{\mu\nu}$ -- $(g^{\mu\nu} \!+\! dg^{\mu\nu}) ]$ 
in an infinitesimal vicinity of the point $x^\al$. 
Obviously, this description is very different from 
the conventional quantum field theoretic one and 
its physical  
significance 
remains to be  
investigated. 
Note, however, that it opens an intriguing 
possibility to 
approximate 
the ``wave function of the Universe'' 
by the fundamental solution of  equation (3.21).  
This solution is 
indeed  
expected to 
describe  
 an expansion 
of the wave function from the 
primary    
``probability clot'' of the Planck scale   
and assigns a meaning to the ``genesis of the space-time'' 
in the sense that 
 the observation of 
the space-time points 
beyond the primary ``clot'' becomes more and more  
 probable with the spreading  of the wave function.    
The self-consistency encoded in the ``bootstrap condition''  
plays a crucial role in such a ``genesis'': in a sense, the 
wave function itself determines, or ``lays down'', 
the space-time geometry it is to propagate on.    

\newcommand{\toymodel}{ 
{

{\bf 3.} Toy Model 

The analysis of the system of equations () and () 
is left beyond the scope of the present paper. 
Still, we will try to gain some insights 
by considering an oversimplified toy model 
inspired by this system. 

given by 
\beq
.............. 
\eeq

where 

derived from the toy model which corresponds to 
the one-dimensional space-time. In this case let us 
introduce the notation $q$ for the densitized metric 
$h^{\mu\nu}$, and $\hat{p}$ for the operator 
$i\hbar \der / \der q$ corresponding to 
$i\hbar \kappa \der / \der h^{\mu\nu}$. 
Note that in doing so $\sqrt{g}$ corresponds to 
$1/q$ and the vielbein $e_a^\mu$  corresponds to 
$q$. The analogue of the Dirac matrix in this 
case can be taken to be equal 
to the unit.  

.... $\gamma AB = 0$ --> no spinor connection?? 

..... but there IS an analog of the term $\gamma^\mu\theta_\mu\$  !! 

..... is there a remnant of the $\tilde{\theta}$ term??

Now, the 1D toy model following from () () 
can be writen in the form 
\beq
i\hbar \tilde{q}(t) \der_t \Psi 
- \pi i G q \der/ \der q \Psi 
= -  8\pi G N \hbar^2 
\{ q^2 \frac{\der}{\der q} \frac{\der}{\der q} \}_{ord} \Psi 
- \Lambda \Psi  
\Psi 
\eeq
with the ``bootstrap condition'' 
\beq
\tilde{q}(t)^2 
= \int \frac{dq}{q^2} \Psib(q,t) q^2 \Psi(q,t) 
\eeq 
The constant $N$ in () is a ``very large'' number 
replacing $1/(n-1)$ in (). 

....... 

measure $dq$ -- ? 

} 
}

\section{ 
Conclusion   
} 

 We discussed an application of 
quantization of fields 
based on the De Donder-Weyl 
canonical 
theory,   
viewed as  
a manifestly covariant generalization of 
the Hamiltonian formulation 
from mechanics to field theory, 
 to the problem of quantization of gravity.

The analogue of the Hamiltonian formulation 
which 
underlies 
 the present procedure of quantization does not  
need  a distinction 
between 
space and time dimensions 
 for,  it treats  
fields as systems 
{\em varying }  
in space-time 
rather than those {\em evolving } in time. 
The De Donder-Weyl canonical equations (1.1)  
describe this type 
of varying 
in space-and-time.  
As 
a 
result, 
the quantum counterpart of the theory 
is formulated on 
a 
finite dimensional configuration 
space of space-time and field variables, with 
the 
corresponding wave {\em function } $\Psi(x^\mu, y^a)$ 
naturally interpreted as 
the 
probability amplitute of 
a 
field 
to have values  
 in  
the interval [$y - (y\!+\!dy)$] in the vicinity of 
the 
space-time point $x$. 
Correspondingly, all the dependence on 
 a 
space-time location 
is transfered 
from operators to the wave function, 
  corresponding to  
 what could be called the 
``ultra-Schr\"odinger'' 
 picture.   

One of the unsolved problems of the 
present approach to be mentioned 
is its 
connection with 
the contemporary notions of quantum field theory. 
The lack of a proper understanding of this issue 
so far 
has been preventing 
specific applications   
of the present framework 
(see, however, 
\cite{castro} for a recent attempt to apply it 
to quantization of $p$-branes).  
 Nevertheless, 
one can hope   
that the already understood character of 
connections  
 between the De Donder-Weyl canonical theory 
and the standard Hamiltonian formalism 
may help 
us 
to clarify this problem. 
Namely, the De Donder-Weyl formulation 
in a sense 
 is 
an intermediate step 
between the manifestly covariant Lagrangian formulation 
and the 
  canonical Hamiltonian 
 framework 
\cite{sternberg,gotaymulti1}.  
In particular, the standard symplectic form 
and the standard equal-time canonical brackets 
in field theory 
can be 
obtained by the integration 
of the polysymplectic form $\Omega$ 
and the canonical brackets (2.3) 
over 
the Cauchy data surface in the covariant 
configuration space $(x^\mu,y^a)$ \cite{romp98}. Similarly,  
the standard functional derivative 
field theoretic 
Hamilton and Hamilton-Jacobi equations 
 can be deduced 
from the DW Hamiltonian and 
the DW Hamilton-Jacobi 
equations. 
The corresponding 
 derivations 
involve a 
pull-back 
of the quantities of DW formulation to a Cauchy data 
surface 
$\Sigma$ given by  $(y^a = y^a_{in}(\bx), t=t_{in})$ 
and a subsequent integration over it. 
 It is natural to inquire if similar connections can be 
established  
between the 
 elements 
of the present approach to field quantization 
and those of the standard canonical quantization. 

Another possible way to establish a connection  with  
the conventional QFT is to view the Schr\"odinger wave functional 
$\Psi([y(\bx)],t)$ 
\cite{hatfield} 
as a composition of amplitudes given by our 
wave functions $\Psi(y,\bx,t)$. 
In  \cite{lodz98}  we discussed this connection 
in the ultra-local approximation  
 but its extension beyond this approximation  
so far remains problematic. 

The character of 
 the 
relationships between the 
DW formulation 
and the conventional canonical formalism  
allows  us to view the former as 
what could be called 
 a 
{\em pre-canonical}  formalism. 
The term reflects the 
intermediate position of the latter 
between the covariant Lagrangian and the 
 canonical Hamiltonian  levels of description.  
Note that in mechanics $(n=1)$ 
pre-canonical 
description coincides with 
 the 
canonical 
 one 
while in field theory $(n>1)$ they become different. 
Hence the questions naturally arise 
as to what 
would be 
a 
pre-canonical analogue  
of quantization and 
 what  
is 
 the 
physical 
significance of the corresponding  
{\em  pre-canonical quantization}.   
The present paper can be viewed  as an attempt  
to shed some light on these questions.  


The application of the present (pre-canonical) 
framework to gravity immediately poses many questions to which 
no 
final 
answers can be given 
 as 
yet. 
Some of these, such as, e.g., 
(i) 
how the spinor wave function 
can be reconciled with the boson vs. fermion nature of the 
fields we are about to quantize,   
(ii) if 
it can or should be replaced 
by a more general Clifford algebra valued wave function,  
(iii) to which extent  one can trust to 
the prescription (3.19) of quantum 
averaging of operators 
notwithstanding 
the underlying scalar product 
is neither positive definite nor space-time location independent, 
(iv) how to quantize operators more general than those 
entering the pre-canonical brackets (2.3),  
and, at last,  
(v) how to calculate observable quantities of interest 
 in field theory 
using the present 
framework, 
concern rather 
the 
 pre-canonical 
               approach in general and 
we hope to address them in 
future publications.  
Let us  
 instead 
concentrate here on a few questions related to 
 the 
specific application to  General Relativity.  

One of the severe problems we encountered is 
due to, on the one hand,  
the non-tensorial nature of 
 basic quantities 
  (polymomenta and the DW Hamiltonian)
of the  
 present 
 DW formulation of General Relativity, 
which  is based  in  essence on the truncated Lagrangian density   
 containing no second-order derivatives of the metric, 
 instead of the generally covariant density $\sqrt{q} R$,  
and, on the other hand, 
the tensorial character of operators 
which only can be 
constructed  
as 
quantum counterparts 
 of these quantities.  
  To avoid  the problem  
  we adopted 
the concept of 
quantization in the vicinity of a point 
 and  
 subsequent covariantization. 
This procedure, however,  
 involves 
 \newcommand{\oldtextc} 
{ basic quantities 
of the 
 DW formulation of General Relativity 
on the classical level 
(the polymomenta and the DW Hamiltonian) 
and, on the other hand, 
the tensorial character of operators 
which only can be 
constructed  
as their quantum counterparts. This could necessiate 
the use of a background connection structure  
very similarly to 
some approaches to specifying 
the  gravitational energy-momentum tensor in General Relativity. 
We, however, 
avoided  
direct tackling 
with  this sort of problems 
by  
adopting  
the concept of 
quantization in the vicinity of a point 
 and 
 subsequent covariantization. 
This procedure, however,  
 still  
 requires some 
 } 
 external elements,  
such as the reference or background vielbein field 
$\tilde{e}{}^\mu_a(x)$ 
and the corresponding spinor connection 
$\tilde{\theta}_\mu (x)$,    
which enter  as non-quantized entities into the generalized 
Schr\"odinger equation (3.21). 
To make the theory self-consistent we introduced 
a supplementary ``bootstrap condition'' (3.18) which 
connects the 
bilinear combination of 
vielbein fields  
$\tilde{e}{}^\mu_a(x)$ 
with the quantum 
 mean value 
of the metric. 
By this means the allowable classical geometrical background 
is included into the theory  
in a 
self-consistent with the underlying quantum dynamics  
way. 
 Clearly, 
this point of view is 
much less radical than 
the usual denial of any 
background geometrical structure 
 which, 
in fact,  
is the source of most 
of the conceptual difficulties of 
quantum gravity. However, within the present scheme 
 it 
 seems to offer 
 an alternative to 
 various 
``pre-geometrical'' constructions.

Next,  
 it should be pointed out that the coefficients 
 involving $n$ in (3.13) and (3.22) 
at the present stage cannot be considered   
as reliably established.   
 This is related both to the ordering   
ambiguity and the unreliability of results obtained by 
formal substitution of polymomenta operators (3.12) 
to classical expressions ({for example,  applying 
the similar procedure to the DW Hamiltonian of a  massless 
scalar field $y$ yields the operator 
$-\frac{n}{2} \hbar^2 \kappa^2 \der^2_{yy}$ 
instead of the correct one 
$-\frac{1}{2} \hbar^2 \kappa^2 \der^2_{yy}$} \cite{qs96,bial97}).  
Let us note also, that at this stage it is 
 rather 
difficult to choose  
between the formulation based on the operator 
of DW Hamiltonian $\what{H}$ 
and  that based on the corresponding density 
$\what{{\cal H}}$. In the 
 former  case,  
the generalized Schr\"odinger 
equation is modified as follows: 
$i\hbar\kappa 
\what{\mbox{$ \gamma^\mu \nabla_\mu$}} \Psi = 
\what{{H}} \Psi $ 
which  in general is different from (3.7)  
due to the ordering ambiguity. 
A preliminary consideration of 
 toy one-dimensional models corresponding to the formulations using  
 respectively 
$\what{H}$ and $\what{{\cal H}}$ 
indicates that the latter formulation, which leads to a toy model 
similar 
to that discussed long ago by Klauder \cite{klauder}, reveals 
more interesting 
behavior and thus may be considered as more suitable.  
To present more conclusive results, however,  
an additional analysis, 
possibly based on quantization of more general  
Poisson brackets than 
pre-canonical ones in (2.3),    is required.   
 
Moreover, as we have already 
pointed out,  
the vielbein  formulation of General Relativity  
is 
a more suitable 
starting point for the application of 
pre-canonical quantization 
 to gravity.  
 The corresponding 
 analysis is in progress and 
 we hope to present it  
 elsewhere.  
Needless to add that much work has to be done to 
 make the present  
  approach 
comparable with 
  other 
  developments in quantum theory of gravity.

   \newcommand{\topicstodiscuss}{  
1. the use of non-tensor quantities is  partially due to an attempt 
to use the formalism developed for the first order Lagrangian 
systems to the E-H Lagrangian which included second-order 
derivatives of  metric. --> the use of 
the truncated E-H Lagrangian which is not scalar --> problems. 
 
2. way out --> $R^2$? and the first order formalism 
in dg and dGamma 
???  prelim considerqation does not work. 

3. gauge fixing = choice of coordinates. how to implement? 
clear, via q averaging.

1a. vielbein formulation can only soften these problems or allow 
to look on them from another perspecive (c.f. consideration of 
the gravitational energy-momentum ``tensor'' within the vielbein 
approach). 

} 

    \section*{Acknowledgments} 
I thank Prof. J. Klauder for very useful 
discussions 
  and for drawing my attention to his 
 earlier papers \cite{klauder}.  
 I also thank Prof. A. Borowiec 
for his valuable remarks 
and C. Castro for  stimulating 
conversations and 
 useful comments and suggestions.  
 I gratefully acknowledge the 
Institute of Theoretical Physics of the 
Friedrich Schiller University of  Jena,   
and Prof. A. Wipf   
\, for kind hospitality and excellent 
working conditions which 
enabled me to finish this paper.




\end{document}